\documentclass[conference]{IEEEtran}
\IEEEoverridecommandlockouts
\usepackage{cite}
\usepackage{amsmath,amssymb,amsfonts}
\usepackage{algorithmic}
\usepackage{graphicx}
\usepackage{textcomp}
\usepackage{xcolor}
\usepackage{color}
\usepackage{url}
\usepackage{hyperref}
\usepackage[latin9]{inputenc}
\usepackage{epstopdf}
\usepackage{amsthm}
\usepackage{amsmath}
\usepackage{amsthm}
\usepackage{amssymb}
\usepackage{graphicx}
\usepackage{subfigure}
\usepackage{wrapfig}
\usepackage{algorithmic}
\usepackage{algorithm}
\usepackage{mathrsfs}
\usepackage{float}
\usepackage{mathtools}
\usepackage{tabulary}
\usepackage{booktabs}
\usepackage{caption}
\usepackage{multirow}
\usepackage{array}
\usepackage{makecell}
\usepackage{graphicx}
\usepackage{enumitem}
\usepackage{subcaption}

\usepackage{titlesec}
\titlespacing*{\section}      {0pt}{*1.55}{*0.25}  % ~12pt before, ~5pt after
\titlespacing*{\subsection}   {0pt}{*0.5}{*0.2}   % ~10pt before, ~4pt after
\titlespacing*{\subsubsection}{0pt}{*0.35}{*0.1}   % ~8pt before, minimal after

\def\BibTeX{{\rm B\kern-.05em{\sc i\kern-.025em b}\kern-.08em
    T\kern-.1667em\lower.7ex\hbox{E}\kern-.125emX}}

\begin{document}

%\title{On the Impact of Device Temperature on RF Fingerprinting Identification: A Novel Temperature-Aware 
%Approach
% {\footnotesize \textsuperscript{*}Note: Sub-titles are not captured for https://ieeexplore.ieee.org  and
% should not be used}
%\thanks{This work is supported in part by NSF Award 2350214.}
%}

%\title{On the Impact of Device Temperature on RF Fingerprinting Identification: A Temperature-Aware Approach}

%\title{On the Impact of Device Temperature on RFFP Identification: A Temperature-Aware Approach}

 \title{Characterizing and Mitigating the Effects of Device Temperature on RF Fingerprinting Accuracy}

\author{\IEEEauthorblockN{Haytham Albousayri}
\IEEEauthorblockA{
% \textit{Department of EECS} \\
\textit{Oregon State University}\\
albousah@oregonstate.edu}
\and
\IEEEauthorblockN{Bechir Hamdaoui}
\IEEEauthorblockA{
% \textit{Department of EECS} \\
\textit{Oregon State University}\\
hamdaoui@oregonstate.edu}
\thanks{This work is supported in part by NSF Award No. 2350214.}}

\maketitle

\begin{abstract}
Radio Frequency Fingerprinting (RFFP) has emerged as a promising approach for device authentication by exploiting hardware-specific impairments embedded in transmitted signals. Yet existing methods largely overlook a major drawback: RFFP sensitivity to temperature--a critical factor influenced by both internal and environmental conditions--which can significantly alter device signatures and degrade classification performance. In this paper, we propose a novel temperature-aware RFFP framework that explicitly incorporates device temperature information into the learning process to improve robustness and generalization.
We evaluate the proposed method on a real-world Bluetooth Low Energy (BLE) dataset collected across multiple devices and environmental conditions. Experimental results demonstrate that temperature-aware modeling consistently outperforms other temperature mitigation baselines, achieving significant improvements in classification accuracy, particularly under unseen temperature and environmental conditions.

% Radio Frequency Fingerprinting (RFFP) has emerged as a promising approach for device authentication by exploiting hardware-specific impairments embedded in transmitted signals. However, existing methods often overlook temperature effects--a critical factor influenced by both internal and environmental conditions--that can significantly alter device signatures and degrade classification performance. In this paper, we propose a novel temperature-aware RFFP framework that explicitly incorporates device temperature information into the learning process to improve robustness and generalization.
% We evaluate the proposed method on a real-world Bluetooth Low Energy (BLE) dataset collected across multiple devices and environmental conditions. Experimental results demonstrate that temperature-aware modeling consistently outperforms other temperature mitigation baselines, achieving significant improvements in classification accuracy, particularly under unseen temperature and environmental conditions. 
\end{abstract}


\section{Introduction}
\label{sec:introduction}
The field of Deep Learning (DL)-based Radio Frequency Fingerprinting (RFFP) has been growing as it brings a data-driven approach for learning device-specific hardware signatures directly from raw RF signals, enabling accurate identification of wireless transmitters without relying on upper-layers security mechanisms. By leveraging the ability of DL models to capture signal distortions caused by hardware impairments, RFFP systems have demonstrated strong performance across various wireless technologies and deployment scenarios. 

Over the years, studies have mainly focused on improving the robustness of these systems against unseen deployment setups, including deploying pretrained RFFP systems on different unseen wireless channels~\cite{johnson2025domain, yuan2025robust}, different time frames~\cite{elmaghbub2024distinguishable}, and even different carrier frequencies~\cite{albousayri2025neural, albousayri2025bluetooth}. 
% Overall, these studies have succeeded in modeling and presenting creative solutions. 
In this study, we shift our attention to another significant degradation factor in RFFP systems, namely the device temperature. 

\begin{figure}
    \centering
    % \hspace{-8pt}
    \subfigure[Identification Accuracy]
    {\includegraphics[width=0.665\linewidth]{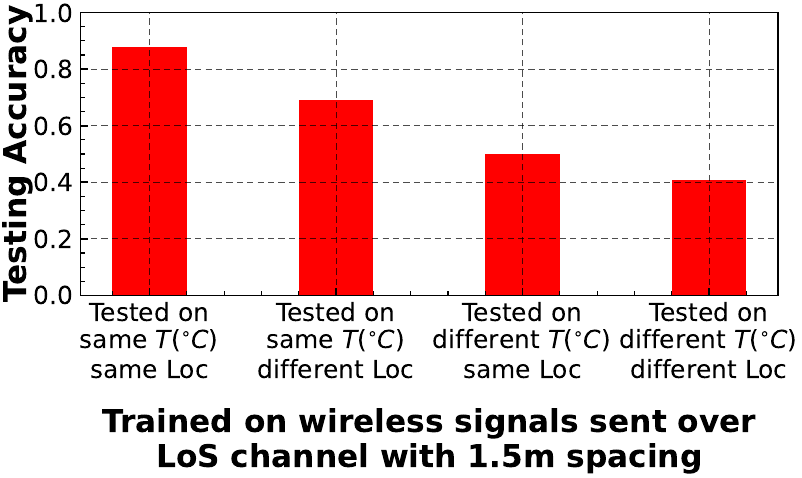}\label{subfig:AccuracyMotivation}}
    \hspace{-8pt}
    \subfigure[CFO vs Temperature]{\includegraphics[width=0.34\linewidth]{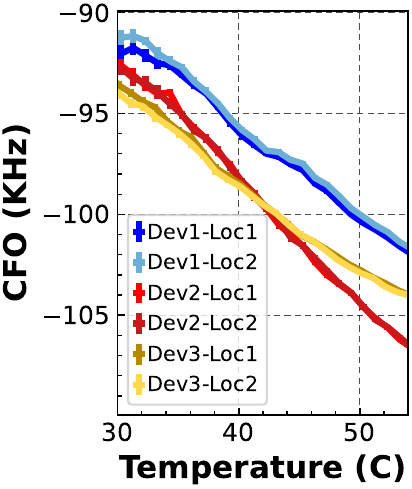}\label{subfig:CFOsVsTemp}}
    \caption{The impact of varying temperatures and locations on RFFP identification for 12 Bluetooth (BLE) devices.}
    \label{fig:Motivation}
\end{figure}

Beyond the impact of the wireless setup on hardware impairments, the internal device temperature plays a critical role in accurate device identification. For instance, an RFFP system trained to authenticate devices at a specific temperature may fail when the temperature changes, causing a large performance drop. Figure~\ref{subfig:AccuracyMotivation} illustrates the impact of changes in device location and temperature on the performance of an RFFP classifier trained on data collected at a specific location and temperature. It can be observed that, while changing the location (with temperature being held constant) causes only a slight drop in performance, changes in temperature introduce much greater confusion for the classifier, leading to a more significant degradation. More details on this experiment are presented later in Section~\ref{sec:dataset}.
This observed behavior is due to the fact that key impairments---such as the carrier frequency offset (CFO)---are shown to be highly correlated with the device's temperature~\cite{gu2024rf}. This can be seen in Figure~\ref{subfig:CFOsVsTemp}, which plots the estimated CFO values of signals coming from three devices, captured from different locations, and under different temperature values. The figure demonstrates a strong correlation between CFO and temperature, while suggesting that this relationship is location independent.

\subsection{Related Work}
In recent years, significant efforts have focused on improving the robustness and generalization of RFFP systems under
changing deployment conditions~\cite{elmaghbub2021lora, johnson2025domain, hui2025cross, xing2022design, peng2024channel, tang2024causal, ma2025towards, yuan2025robust, elmaghbub2023eps, albousayri2025bluetooth}. These conditions include training the RFFP model on data collected at a specific location, channel, and/or time, and later evaluating its performance under different unseen domains (other location, channel, and/or time). However, the impact of device temperature on RFFP performance, where the models are trained on data collected at certain temperatures and later deployed and tested under different device temperature conditions, has been barely explored. 

Among the few studies that have investigated device-temperature effects is the work of Elmaghbub et al.~\cite{elmaghbub2024no}, who examined the impact of device warm-up on RFFP performance and demonstrated that identification accuracy can be highly sensitive to hardware impairment changes occurring during the device stabilization phase.
Although temperature was not explicitly measured, their study captured device signals over a 30-minute interval to observe the temporal evolution of RFFP impairments, which effectively parallels tracking their variation as the device temperature rises during operation. Their results showed that classifiers trained on data collected during the warm-up phase (the initial few minutes) failed to accurately identify devices when tested on data acquired at later times, highlighting the combined impact of temperature variation and temporal dynamics.
More recently, the authors in~\cite{yilmaz2022effect, nieves2026measurements} studied the sensitivity of RFFP systems to changes in device temperature. They experimentally confirmed that RFFP systems are susceptible to temperature variations and highlighted significant performance degradation when cross-testing classifiers across different temperature conditions; however, no solutions were proposed to address this issue.

To mitigate the impact of temperature, the authors in~\cite{shen2022towards} proposed removing the residual CFO---a dominant RFFP feature---from the signal prior to classification. Their approach assumes that temperature primarily affects the CFO impairment, thus removing it from the signal yields a more temperature-invariant RFFP representation.
Furthermore, the authors in~\cite{gu2024rf} investigated temperature variations in RFFP systems. In their proposed approach, the transmitters are assumed to report their temperature, and a separate RFFP model is trained for each temperature value. However, the method is inherently limited to a finite set of temperature points, with evaluation performed on only eight discrete integer temperature levels. Because temperature is a continuous variable, this design raises concerns about scalability and generalization, and underscores the need for a temperature-aware framework capable of operating across a continuous range of values.

\subsection{Contributions}
The main contributions of this work are summarized as follows:
\textit{1)} We propose a scalable, temperature-aware RFFP framework that incorporates device temperature as an input feature, enabling robust operation across a continuous range of temperatures, mitigating its impact on classification performance, and directly addressing the limitations of existing methods.
\textit{2)} We evaluate the proposed framework on a real-world dataset spanning multiple devices, locations, and environmental conditions, including both line-of-sight (LoS) and non-line-of-sight (NLoS) transmissions.
\textit{3)} Experimental results demonstrate that our approach achieves over 60\% improvement in identification accuracy compared to existing baselines across challenging and unseen temperature and environmental conditions. 
\textit{4)} We have made our dataset publicly available at \url{https://research.engr.oregonstate.edu/hamdaoui/datasets} for reproducibility and benchmarking.
% \subsection{Contribution}
% To address these gaps, we propose a scalable, temperature-aware RFFP framework that operates effectively over a continuous range of temperatures. The proposed approach incorporates device temperature as an input feature to the RFFP model, thereby mitigating the impact of device temperature on RFFP systems. Our experimental results show that the proposed approach achieves over 60\% improvement in the identification accuracy compared to existing baselines under different challenging environments, including both line-of-sight (LoS) and non-line-of-sight (NLoS) transmissions. 

%     Our dataset is made publicly available for the community.
 
The remainder of this paper is organized as follows: Sections~\ref{sec:hw-impairments} highlights the key hardware impairments. Section~\ref{sec:system-model} introduces the proposed system model. 
Section~\ref{sec:method} presents our temperature-aware RFFP framework. Section~\ref{sec:dataset} describes our experimental testbed and dataset collection. Section~\ref{sec:results} presents our results. Finally, Section~\ref{sec:conclusion} concludes the paper.

\section{Hardware Impairments Background}
\label{sec:hw-impairments}

Because RFFP systems rely primarily on the extraction and exploitation of hardware-induced imperfections in transmitting devices, we begin this section by providing the essential background needed to understand these impairments. 
In practice, each device exhibits unique hardware impairments and distortions introduced during manufacturing and operation. These hardware impairments, arising from imperfections in RF hardware components such as local oscillators, mixers, and power amplifiers, cause transmitted signals to deviate from their ideal characteristics. Although traditionally considered undesirable in communication systems, these impairments create distinctive signatures that can be leveraged to differentiate between devices. These include, but are not limited to:
\begin{itemize}[leftmargin=*]
\item \textbf{Carrier Frequency Offset and Phase  Noise:}
Carrier frequency offset (CFO) arises from frequency mismatches between the transmitter's and receiver's local oscillators, resulting in a frequency shift and a time-varying (progressive) phase rotation. In contrast, phase noise stems from phase mismatches between the oscillators and appears as a random induced phase shift. Both impairments arise from inherent imperfections in the local oscillator and phase-locked loop circuitry.
% \item \textbf{Maximum Frequency Deviation Offset:} 
% This represents the error in the nominal maximum frequency deviation, which can be defined as the difference between the maximum positive frequency and the central frequency in GFSK modulation.
\item \textbf{I/Q Amplitude and Phase Imbalances:} 
These capture the imbalance arising when the in-phase (I) and quadrature (Q) components of the transmitter and receiver are mismatched in amplitude and/or phase, most often due to imperfections in mixers and local oscillators. These typically result in constellation distortion and image frequency leakage.

\item \textbf{I/Q DC Offsets:} This occurs when the I/Q signal origin shifts from zero due to mixer leakage or DAC imperfections, producing a constant DC component and carrier feedthrough in the transmitted signal.

\item \textbf{Power Amplifier Non-linearity:} 
This distortion arises when the transmitter power amplifier operates outside its ideal linear region, causing amplitude and phase distortions, especially at higher input power levels. Such non-linear behavior can introduce spectral regrowth, waveform distortion, and undesired harmonics, thereby altering the transmitted signal characteristics in a device-dependent manner.
\end{itemize}

% Mathematically, the set of underlying hardware impairments associated with the $k$-th device is denoted as $\mathbf{\Theta}_k$.
Throughout, the set of these hardware impairments associcated with device $k$ will be denoted by the vector $\mathbf{\Theta}_k$. 
Collectively, these impairments uniquely distort a device's transmitted RF signal, allowing a receiver to observe distinctive fingerprints in captured transmissions that can ultimately be used for device identification.
Therefore, a capable RFFP system can effectively map the observed distortions in the captured signal to the corresponding device identity, without explicitly estimating $\mathbf{\Theta}_k$. This study further shows that, while these impairments are highly distinctive across devices, they are also strongly influenced by the device's internal temperature, as illustrated in Figure~\ref{subfig:CFOsVsTemp}, thereby making the captured signals temperature-dependent.

% these effects can be modeled as $\mathbf{\Theta}_k(T)$, where $T$ denotes the device temperature in $^\circ{C}$, and hence making the captured signals from these devices temperature-dependent. 

% exploit such distortion  as unique fingerprints for device identification.
% with each device exhibiting a distinct statistical distribution, 
% allowing them to be exploited as unique fingerprints for device identification.
 
% This study shows that $\mathbf{\Theta}_k$, while very unique for each device, is highly dependent on the internal temperature, as illustrated in Figure~\ref{subfig:CFOsVsTemp}, and can therefore be modeled as $\mathbf{\Theta}_k(T)$, where $T$ denotes the device temperature in $^\circ\text{C}$.

\section{RFFP System Model}
\label{sec:system-model}
\begin{figure}
    \centering
    \includegraphics[width=0.8\linewidth]{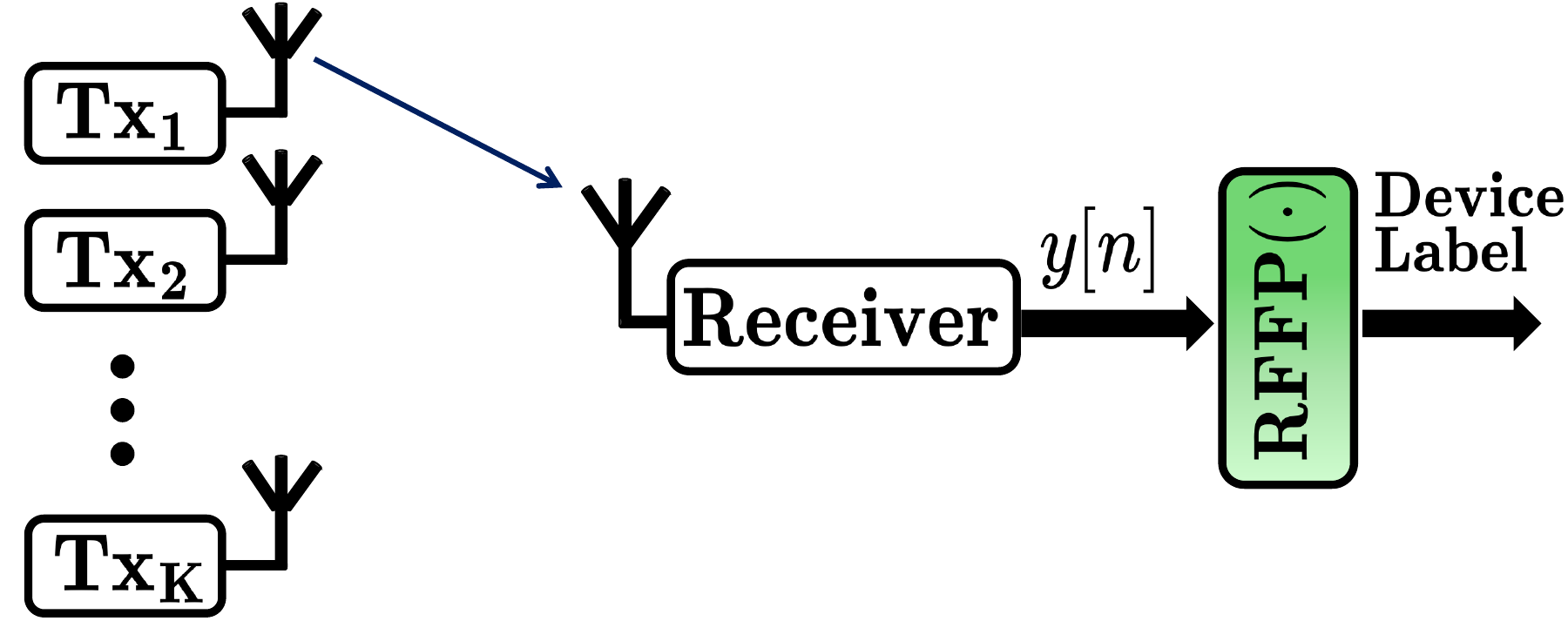}
    \caption{Overview of our RFFP system.}
    \label{fig:system_overview}
\end{figure}
% Given a transmitter that sends a 
As shown in Figure~\ref{fig:system_overview}, our proposed RFFP systems consists of $K$ transmitting devices, a receiver (Rx), and an $\mathbf{RFFP}(\cdot)$ function serving as the authenticator. 
Each transmitting device $k$ is associated with a unique impairments vector $\mathbf{\Theta}_k$ whose impact on the device's transmitted signal 
%Collectively, these impairments distort ideal RF signals and their impact 
can be modeled as~\cite{albousayri2025neural, albousayri2026replicating} $\mathcal{F}_{\text{RFFP}}(\cdot, \mathbf{\Theta}_k)$. Therefore, the baseband signal received from the $k$-th device can be expressed as:
\begin{equation}
    y_k[n]= h[n] * \mathcal{F}_{\text{RFFP}}(x[n], \mathbf{\Theta}_k) + \eta[n],
    \label{eq:system-model}
\end{equation}
where $h[n]$ is the channel response, $x[n]$ is the transmitted (ideal) baseband signal, $\eta[n]$ is an additive white gaussian noise term, and $*$ represents the convolution process. 

Once the receiver captures the signal, it is then fed into the $\mathbf{RFFP}(\cdot)$ block, which predicts the identity of the transmitting device solely based on $y_k[n]$; that is,
\[
k = \mathbf{RFFP}(y_k[n]).
\]
% In general, the decision boundaries of this 
In realistic setups, $y_k[n]$ is influenced not only by the transmitter-specific impairments $\mathbf{\Theta}_k$, but also by the channel conditions $h[n]$ and the receiver's hardware impairments. These additional factors can cause further distortions on the observed signal, potentially leading to significant increases in the identification errors. 

Over the years, researchers have focused on building the $\mathbf{RFFP}(\cdot)$ block using deep learning, primarily to improve its consistency and robustness under various domain changes, including changes in locations, receivers, channel, and/or time. However, limited attention has been given to the impact of device temperature.
To address this research gap, in this study, we investigate the impact of a device's internal temperature on the behavior of the hardware impairments $\mathbf{\Theta}_k$---and consequently on the received signal $y_k[n]$---and propose an RFFP framework that incorporates the device's temperature as an additional feature to the RFFP model, thereby improving the robustness of RFFP systems under temperature variations.

\section{Temperature-Aware RFFP Identification}
\label{sec:method}

Prior studies have focused on designing DL-RFFP classification models that only use the captured signal $y_k[n]$ for predicting device $k$. As discussed earlier, while these approaches work under minor temperature variations, they fail under realistic (large) temperature changes. 

To address this limitation, we propose a novel temperature-aware DL-RFFP framework that incorporates the device temperature $T_k$ as an additional feature, i.e., $k = \mathbf{RFFP}([y_k[n]; T_k])$, enabling improved robustness under varying temperature conditions.
Our framework is built with two key design considerations in mind: (i) real-time over-the-air (OTA) temperature sharing; and (ii) temperature-dependent DL-RFFP classifier design, which are described in more detail next.

\subsection{Payload Design and Temperature Measurement}
\begin{figure}
    \centering
    \includegraphics[width=\linewidth]{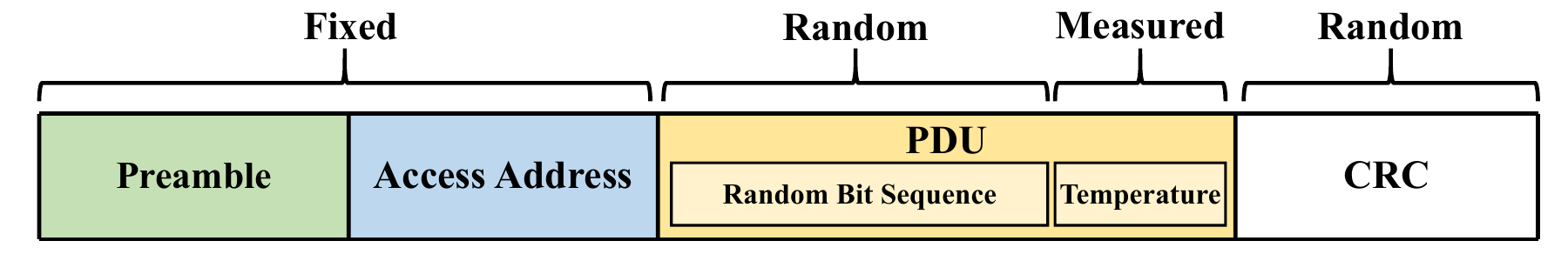}
    \caption{Designed BLE frame.}
    \label{fig:BLE_frame}
\end{figure}

We propose capturing real-time temperature measurements using the device's internal sensor, encoding them in the last two bytes of the payload field, and transmitting them over the air (OTA) to the receiver.
As shown in Figure~\ref{fig:BLE_frame}, all BLE devices were configured to periodically transmit advertisement packets (beacons), each consisting of a fixed preamble and access address, along with a random payload that includes the device's internally measured temperature.

\begin{figure}
    \centering
    \includegraphics[width=\linewidth]{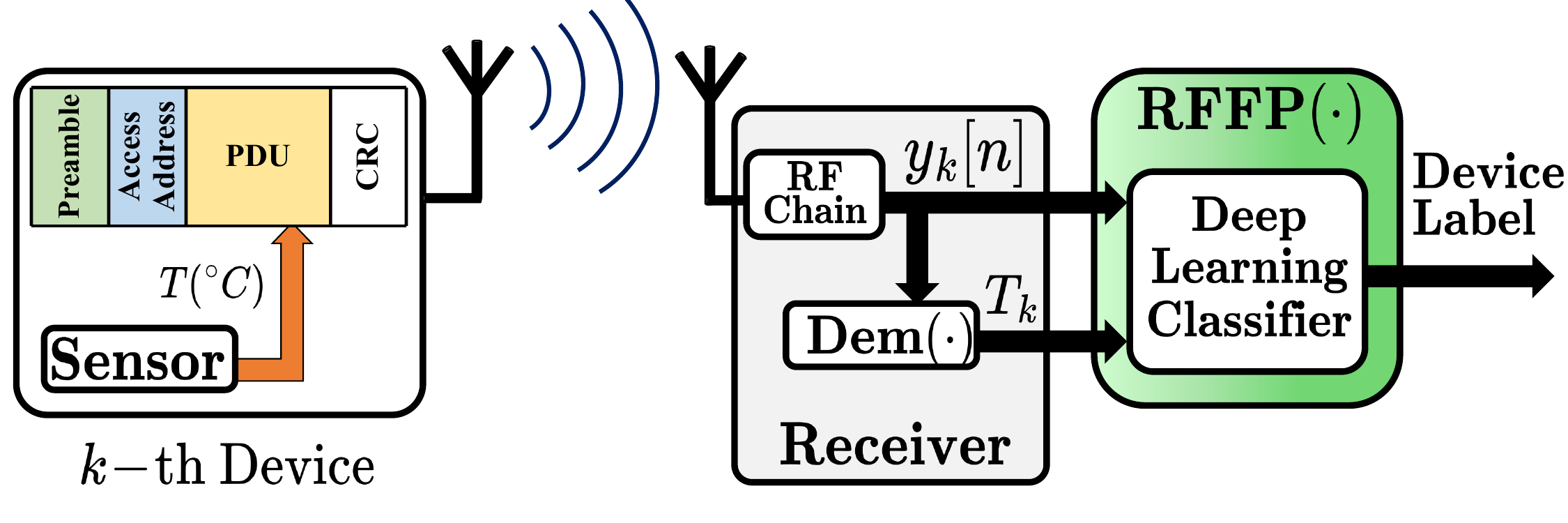}
    \caption{Proposed temperature-aware RFFP framework.}
    \label{fig:payloadOTA}
\end{figure}

Once the receiver captures the PHY-layer OTA frame $y[n]$--where the index $k$ is dropped for simplicity--it is first synchronized and power-normalized. The frame is then demodulated to extract the temperature value from the payload, after which both $y[n]$ and the associated temperature are provided to the RFFP stage. Figure~\ref{fig:payloadOTA} illustrates the end-to-end temperature-aware RFFP framework.


\subsection{RFFP via Temperature-Aware Deep Learning}
\begin{figure}[t]
    \centering
    \includegraphics[width=\linewidth]{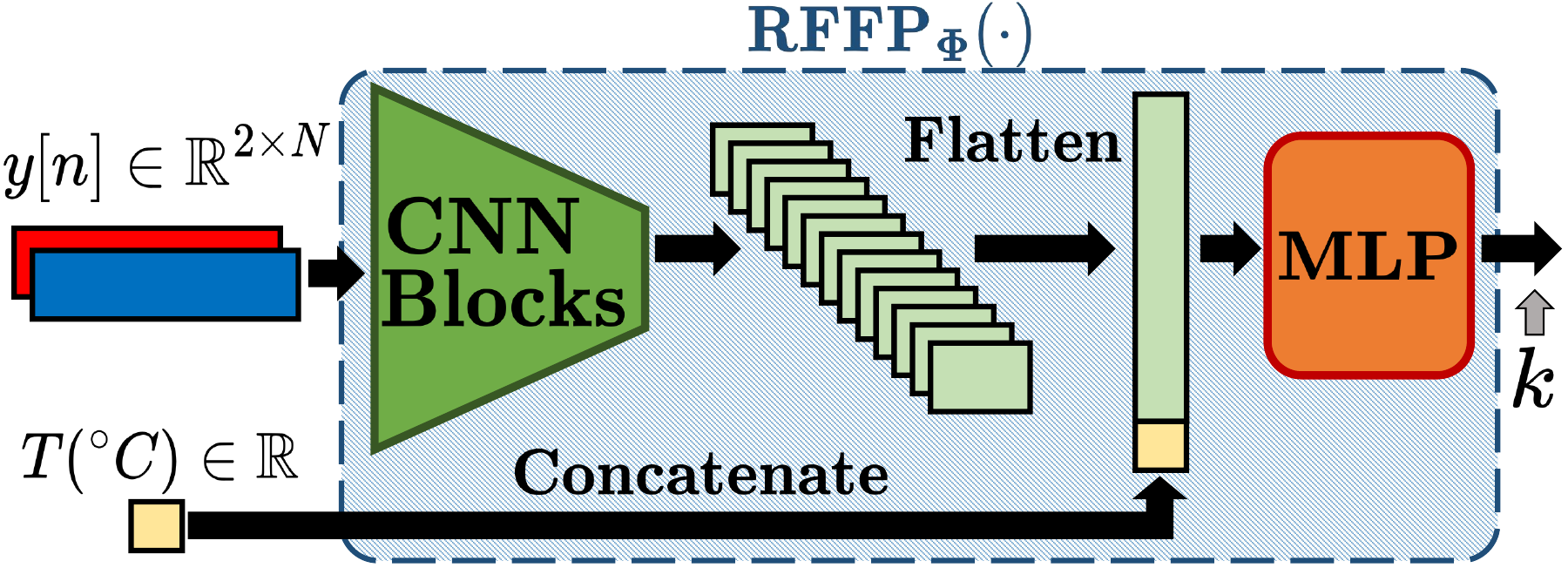}
    \caption{Proposed Training Pipeline.}
    \label{fig:CNN}
\end{figure}

As Convolutional Neural Networks (CNNs) have shown strong performance in domain adaptation for RFFP tasks~\cite{johnson2025domain}--that is when the training and testing setups are mismatched--we leveraged this architecture to perform the classification task. 

Figure~\ref{fig:CNN} shows the proposed temperature-aware classifier, where a CNN encoder processes $y[n]$ to produce feature embeddings. These embeddings are then flattened and concatenated with the temperature measurement $T$. The resulting vector is fed into a multi-layer perceptron (MLP) that outputs the probability distribution over device labels. Including temperature as an additional feature allows the model to account for temperature-induced variations in hardware impairments, enabling it to capture the trajectory of feature changes rather than relying solely on static representations.
% The architecture used for the CNN encoder consists of five convolutional blocks with filter sizes $\{32, 64, 128, 256, 512\}$ and kernel sizes $\{48, 12, 6, 3, 3\}$. Each block applies a 1D convolution with padding set to half the kernel size, followed by a LeakyReLU activation, batch normalization, and max-pooling with kernel size 2 to reduce the temporal resolution. This design allows the shallow layers with larger kernels to capture broad temporal dependencies, while deeper layers with smaller kernels emphasize fine device-specific features.
% The number of neurons used for the MLP stage are $\{1024, 512, K\}$ ($K$ is the number of classes). ReLU activations follow the first two layers, and dropout with rates ${0.4}$ and $0.3$ is applied after the 1024- and 512-neuron layers, respectively.
More details on the proposed DL architecture are presented in Section~\ref{subsec:DLresults}.

\section{Experimental Setup and Dataset Collection}
\label{sec:dataset}
\begin{figure}[t]
    \centering
    \hspace{-8pt}
    \begin{minipage}[b]{0.2\linewidth}
            \centering
            \subfigure[ESP32C3]{
                \includegraphics[width=\linewidth]{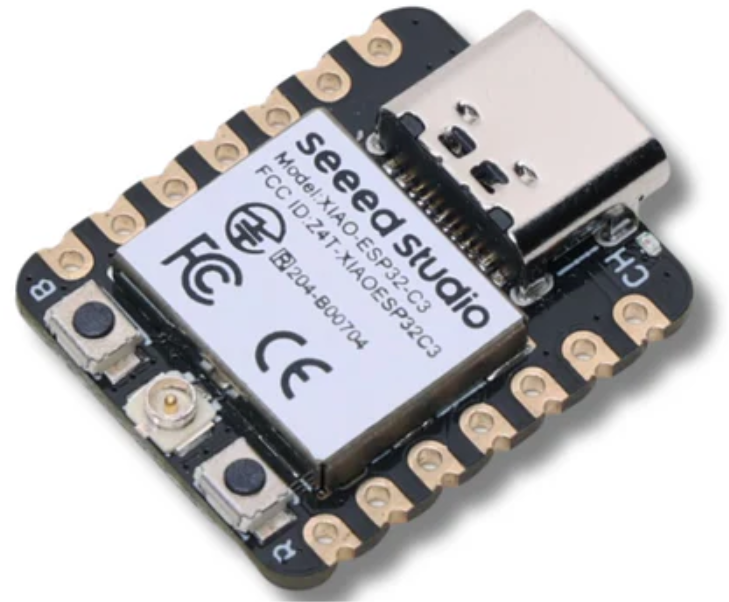}}\\[0.5ex]
                \vspace{-13pt}
            \subfigure[USRP B210]{
                \includegraphics[width=\linewidth]{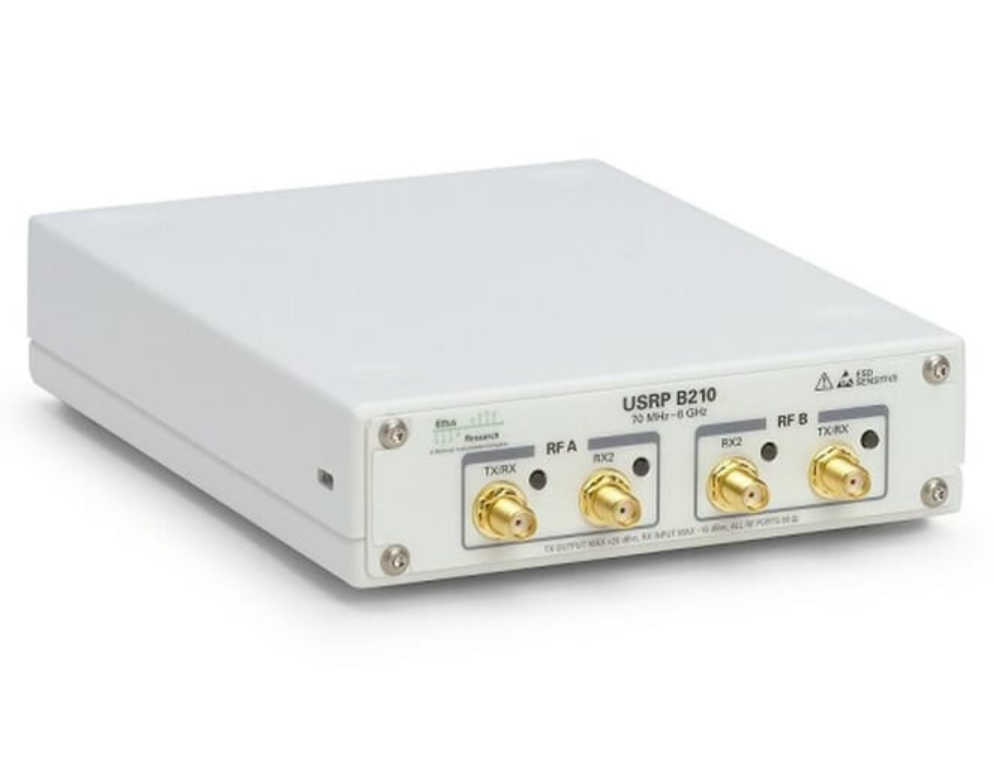}}
    \end{minipage}
    % \hspace{-4pt}
    {\centering
    \begin{minipage}[b]{0.25\linewidth}
        \centering
        \subfigure[Loc1]{
            \includegraphics[width=\linewidth]{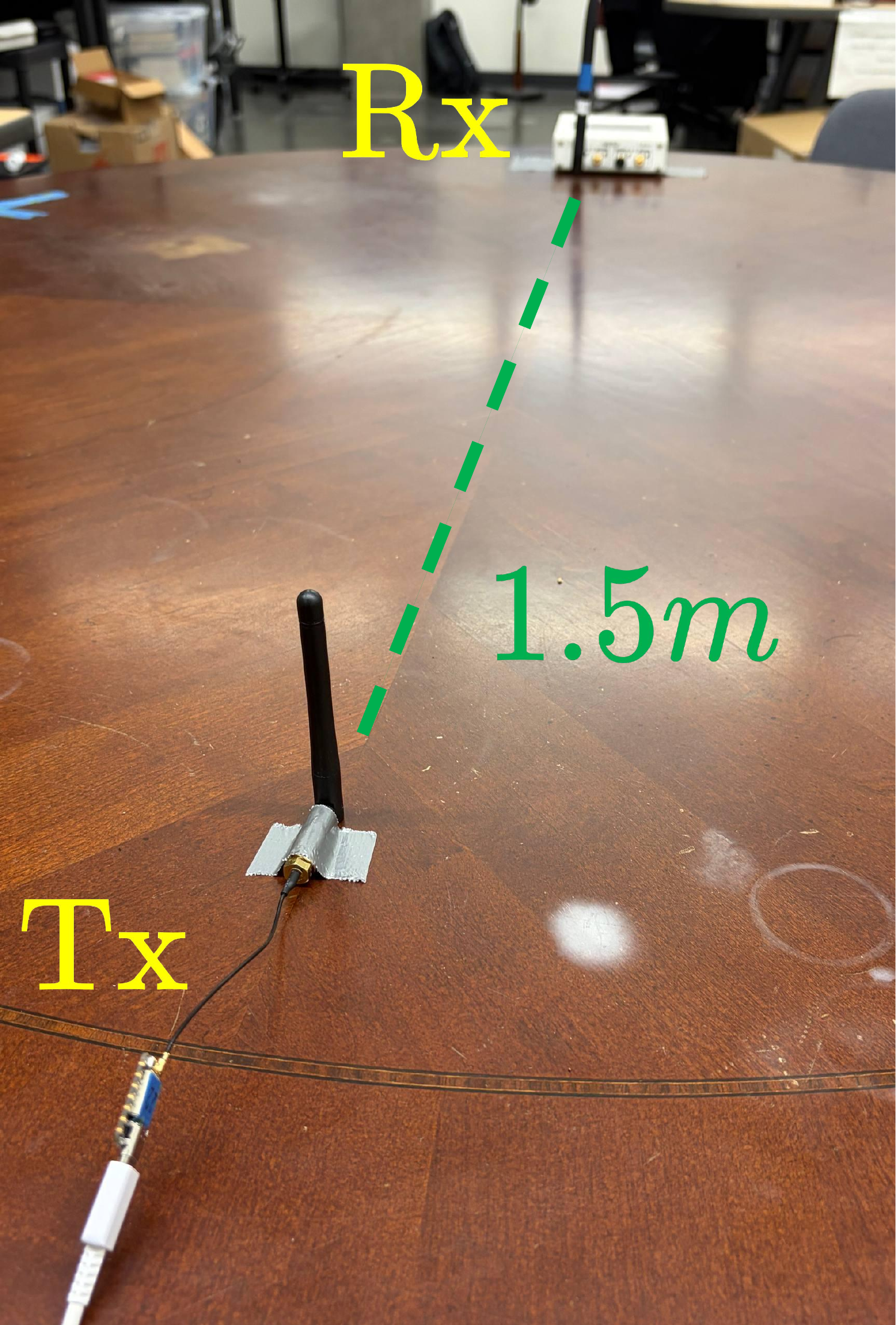}}
    \end{minipage}}
    \hspace{-4pt}
    {\centering
    \begin{minipage}[b]{0.25\linewidth}
        \centering
        \subfigure[Loc2]{
            \includegraphics[width=\linewidth]{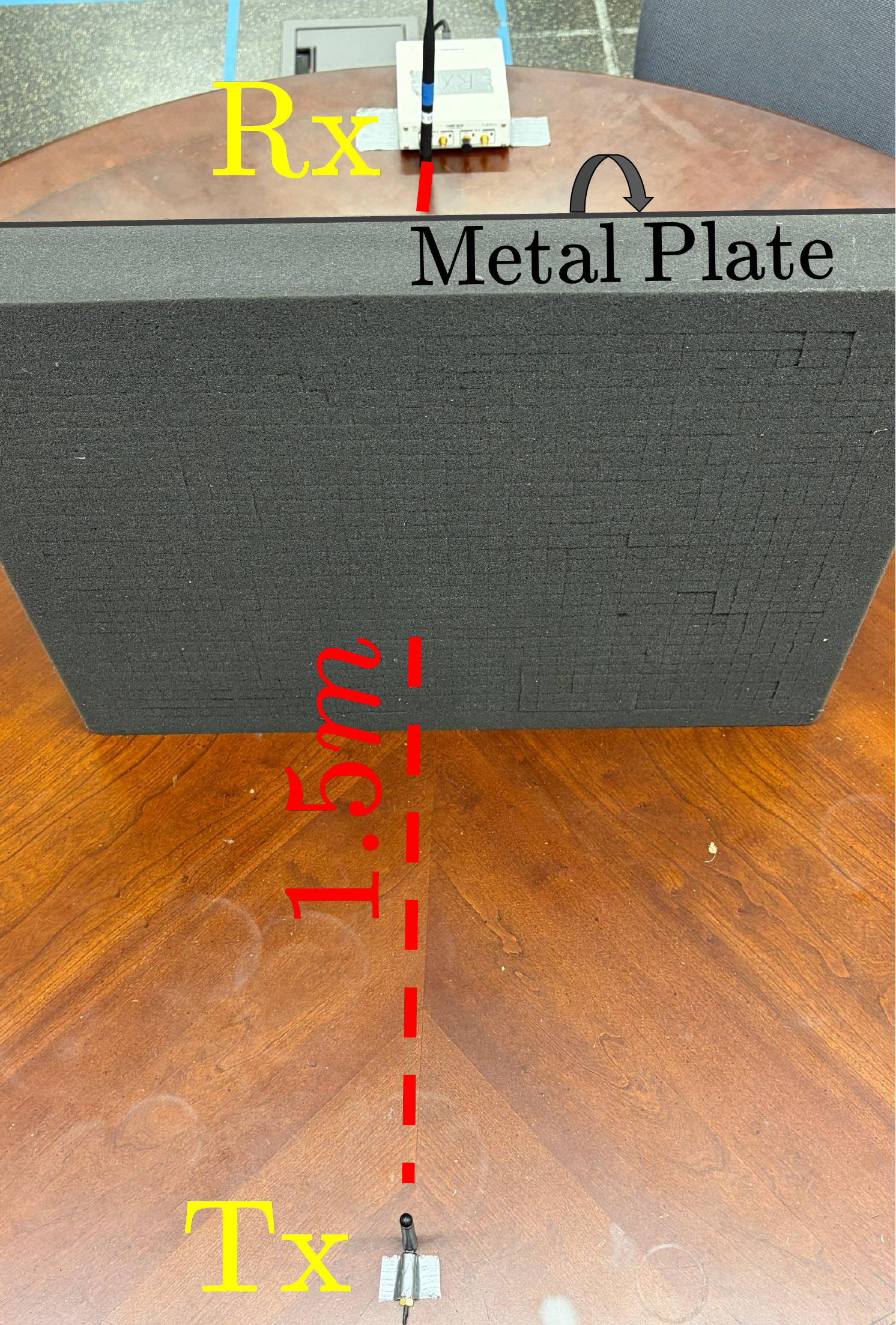}}
    \end{minipage}}
    \hspace{-4pt}
    {\begin{minipage}[b]{0.25\linewidth}
        \centering
        \subfigure[Loc3]{
            \includegraphics[width=\linewidth]{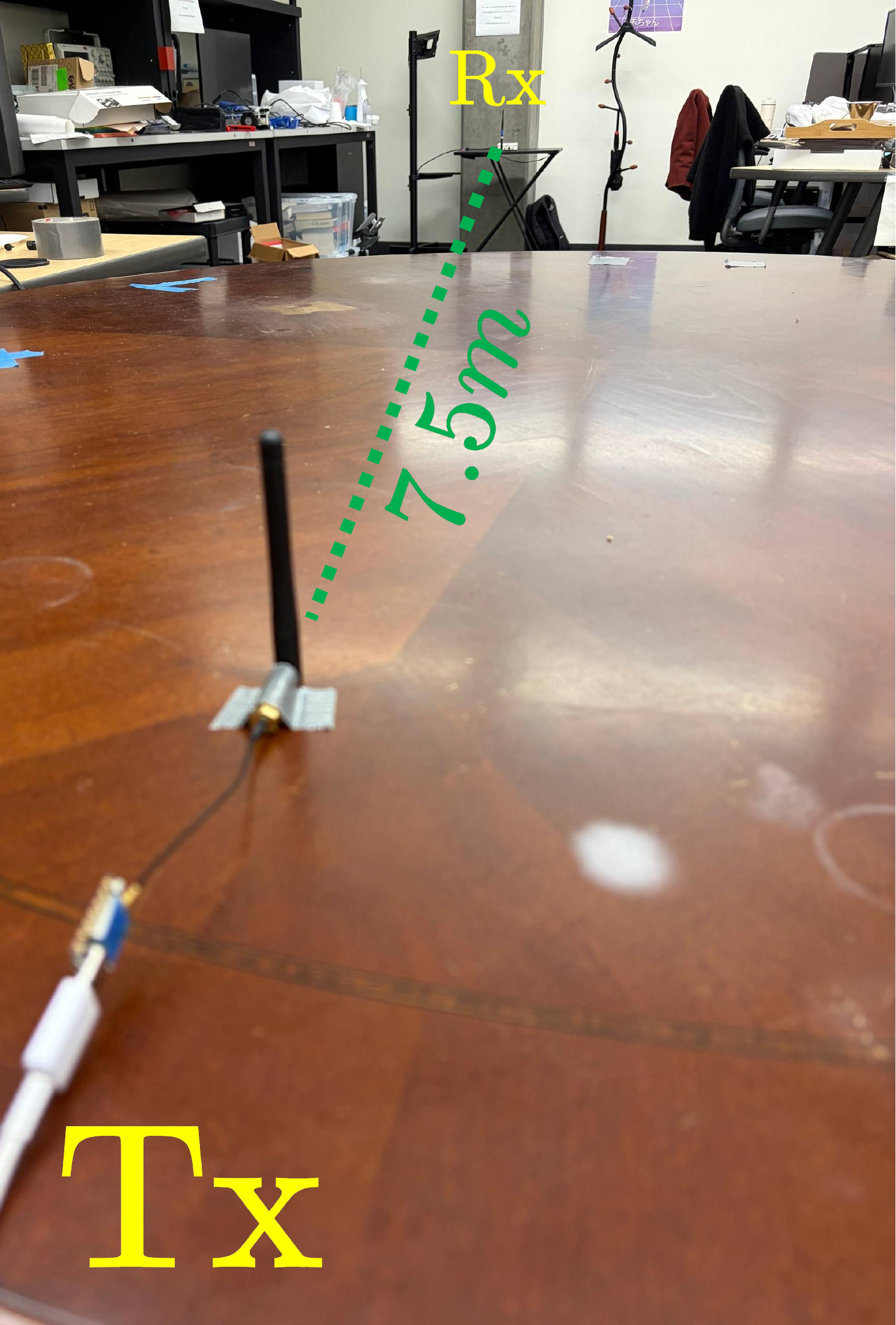}}
    \end{minipage}}
    \caption{Experimental testbed and hardware configuration.}
    \label{fig:Testbed}
\end{figure}

We used 12 Bluetooth Low Energy (BLE) devices from ESP32-C3 Xiao Seeeds chip as the transmitter (Tx) and a USRP B200 as the receiver (Rx), as illustrated in Figures~\ref{fig:Testbed}(a) and (b). 
Three datasets were collected, each corresponding to a distinct Tx-Rx propagation environment:
% Figures~\ref{fig:Testbed}(c-e)
% The devices operate using the uncoded 1M PHY mode over BLE frequency channel 1, centered at 2.406 GHz.
\begin{itemize}[leftmargin=*]
    \item \textbf{Loc1 (Fig.~\ref{fig:Testbed}(c)):} the Tx-Rx path is a 1.5m line-of-sight (LoS) link, yielding an average SNR of 30dB.
    \item \textbf{Loc2 (Fig.~\ref{fig:Testbed}(d)):} a metal conductor is placed between Tx-Rx, yielding a non-line-of-sight (NLoS) link with 1.5m separation and resulting in an average SNR of 20dB.
    \item \textbf{Loc3 (Fig.~\ref{fig:Testbed}(e)):} the Tx-Rx path is a LoS link with 7.5m separation in a rich fading environment, yielding an average SNR of 18dB.
\end{itemize}

%\begin{figure}
%    \centering
%    \subfigure[Training Window]{\includegraphics[width=0.48\linewidth]{figures/TempvsTime_TrainingOfficial.pdf}\label{subfig:TempVsTime}}
    %
%    \subfigure[DL baseline comparision]{\includegraphics[width=0.48\linewidth]{figures/AccuracyVsTime_Official.pdf}\label{subfig:DL_AccVsTime}}
%    \caption{The impact of temperature on RFFP performance: (a) device temperature evolution during hardware warm-up period; (b) RFFP classification accuracy when training at first three minutes but testing at different later times.}
%    \label{fig:WarmUp}
%\end{figure}

\begin{figure}
    \centering
    \includegraphics[width=0.9\linewidth]{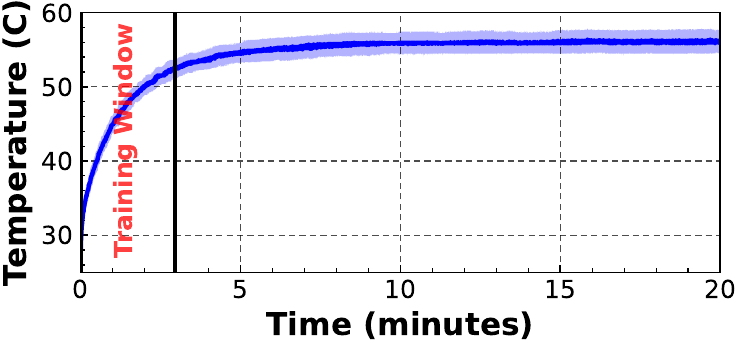}
    \caption{Device temperature evolution during the warm-up period of device hardware.}
    \label{subfig:TempVsTime}
\end{figure}

At the receiver (Rx) side, I/Q samples are continuously captured at a rate of 6 MS/s while each transmitter (Tx) remains powered on and transmits for a duration of 20 minutes. 
Figure~\ref{subfig:TempVsTime} shows the average temperature of all Tx devices during the first 20-minute period. We can observe that the devices go through a warm-up period within the first 5 minutes where the temperature goes from around $30^\circ C$ to $56^\circ C$ and later enters a steady state period. 

Moreover, each complex frame was extracted, carefully synchronized and power normalized. Only the first $N=256$ I/Q time samples are used as a $(2,N)$ real valued tensor, where the first dimension of size 2 is for the I and Q components. This choice of $N=256$ ensures the use of fixed sequences---namely the preamble and part of the access address fields---across all devices.
Overall, the first 20 minutes produce around 33,000 I/Q frames for every device, shaped as $(33,000, 2, 256)$. We only use the first 3 minutes as our training window (around 5000 frames)---illustrated in Figure~\ref{subfig:TempVsTime}. A split of $70\%$, $10\%$ and $20\%$ for training, validation and testing, respectively, is utilized. During testing, we used a 1-minute sliding window (around 1,650 frames) to evaluate the pre-trained classifiers. This window was shifted in 1-minute increments to cover the entire 20-minute recording.

% the pre-trained classifiers were evaluated on the entire 20-minute period in minute-by-minute steps, where in each minute only the unseen testing set was used.
% testing sets, that is, sampled from both the first 5,000 and the remaining 28,000 I/Q frames.

Identification (Classification) accuracy was used as our evaluation metric, and each reported value represents the average and standard deviation of 10 independent seeds.

\section{Experimental Evaluation}
\label{sec:results}
In this section, we evaluate and demonstrate the effectiveness of the proposed temperature-aware RFFP 
approach using our datasets presented in Section~\ref{sec:dataset}.

The CNN-DL architecture used for the encoder block consists of five convolutional blocks with filter sizes $\{32, 64, 128, 256, 384\}$ and kernel sizes $\{48, 16, 8, 3, 3\}$. Each block applies a 1D convolution with padding set to half the kernel size, followed by a LeakyReLU activation, batch normalization, and max-pooling with kernel size 2 to reduce the temporal resolution. This design allows the shallow layers with larger kernels to capture broad temporal dependencies, while deeper layers with smaller kernels emphasize fine device-specific features.
The number of neurons used for the MLP stage are $\{1024, 512, K\}$ ($K$ is the number of classes). ReLU activations follow the first two layers, and dropout with rates ${0.4}$ and $0.3$ is applied after the 1024- and 512-neuron layers, respectively.
The total number of parameters of this configuration is ${4,174,892}$.

\subsection{The Impact of DL Model on RFFP Performance}
\label{subsec:DLresults}

To justify the choice of the proposed CNN architecture, we begin by comparing it with three other DL baselines for the same RFFP classification task. Specifically, we compared the CNN architecture with the following DL architectural models:
\begin{itemize}[leftmargin=*]
    \item \textbf{Transformer~\cite{ahmed2023transformers} [\textit{4,301,486 parameters}]}: A state-of-the-art model for sequential processing tasks, leveraging self-attention mechanisms to capture long-range dependencies and global context efficiently.
    \item \textbf{Gated Recurrent Unit (GRU)~\cite{hewamalage2021recurrent} [\textit{4,235,758 parameters}]}: A recurrent neural network architecture designed for sequential data, using gating mechanisms to capture temporal dependencies while mitigating vanishing gradient issues.
    \item \textbf{Long Short-Term Memory (LSTM)~\cite{al2021deeplora} [\textit{4,238,958 parameters}]}: A type of recurrent neural network that uses input, forget, and output gates to better model long-term dependencies, making it more expressive than GRU at the cost of increased complexity and parameters.
\end{itemize}

% \begin{itemize}
%     \item \textbf{Transformer [4,301,486 params]}: A state of the art model for sequential processing tasks.
%     \item \textbf{Gated Recurrent Unit (GRU) [4,235,758 params]}: 
% \end{itemize}
% namely: \textbf{Transformer}, \textbf{Gated Recurrent Unit (GRU)} and \textbf{Long Short Term Memory (LSTM)}. For fair comparison, the optimal choice of layers/parameters were made such that the total number of parameters are roughly the same. More specifically, the resulting size of these models are 4,301,486, 4,235,758 and 4,238,958 respectively. 

For a fair comparison, the architectural models were designed such that the total number of parameters is approximately the same across all models. To evaluate the sensitivity of these models to temperature variations, we trained them using the proposed approach on captured frames from the first 3 minutes, during which the temperature ranges from $30^\circ C$ to $52^\circ C$. The models are then tested on signals collected at different times, allowing us to assess the effects of both temporal and unseen temperatures.

% To evaluate the sensitivity of these models to temperature variations, we trained them using our proposed approach to identify the transmitters using the captured frames within the first 3 minutes. These frames experience temperature that goes from $30^\circ C$ to $52^\circ C$. We later test our classifier on later collected signals, hence investigating both the temporal and unseen temperature impacts. 

\begin{figure}
    \centering
    \includegraphics[width=0.8\linewidth]{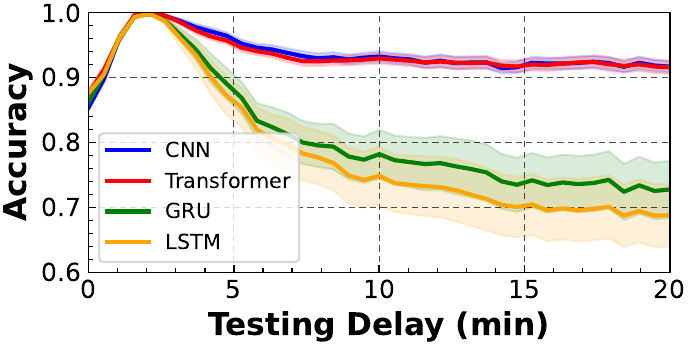}
    \caption{RFFP classification accuracy when training at first three minutes but testing at different later times.}
    \label{subfig:DL_AccVsTime}
\end{figure}

Figure~\ref{subfig:DL_AccVsTime} plots the testing accuracy---averaged across all datasets from the 3 locations---for the studied DL-RFFP baselines as the time/temperature at which the testing data is collected increases. Here, all DL-RFFP models are trained using data collected during first 3 minutes. 
%
% The average accuracy when cross training and testing over all the datasets from the 3 different locations is presented. 
Overall, the figure highlights that \textbf{CNN} and \textbf{Transformer} achieve relatively similar, but significantly better device identification accuracy than \textbf{GRU} and \textbf{LSTM} models. 
Furthermore, Table~\ref{tab:Timecomparison} summarizes the training time and inference time for these DL-RFFP baselines, highlighting the efficiency of \textbf{CNN} compared to all the others, and suggesting that it is best suited for RFFP tasks. 

Therefore, for the remainder of the evaluation, we adopt the CNN architecture in our proposed Temperature-Aware RF fingerprinting identification framework.

% The results also highlights that, while a monotonic decrease is observed, the CNN and Transformer architectures maintain a very high classification accuracy even over large unseen temperature, which is made only possible thanks to our temperature aware RFFP framework as will be presented in Subsection~\ref{subsec:RFFP_Domains}.

\begin{table}
    \centering
    \caption{Execution Time (mean $\pm$ std) comparison of baselines and proposed technique (in milliseconds).}
    \label{tab:Timecomparison}
    \begin{tabular}{l|cc}
        \toprule
        \midrule
        \textbf{Baseline} & \textbf{Training Time (per step)} & \textbf{Inference Time} \\
        \midrule
        {CNN} & $\bf{2.7127 \pm 0.169}$ & $\bf{0.4675 \pm 0.0626}$   \\
        \midrule
        {Transformer} & $5.3691 \pm 0.0421$ & $0.5438 \pm 0.0759$  \\
        \midrule
        {GRU} & $2.9434 \pm 0.0938$ & $0.6375 \pm 0.0284$ \\
        \midrule
        {LSTM} & $2.9624 \pm 0.1656$  & $0.7479 \pm 0.0432$\\
        \midrule
        \bottomrule
    \end{tabular}
\end{table}

% Furthermore, Table~\ref{tab:Timecomparison} summarizes the training time and inference time for these DL-RFFP baselines, highlighting the superiority of CNN over all the others. 

\subsection{RFFP Identification Accuracy}
\label{subsec:RFFP_Domains}

After demonstrating the robustness and justifying the choice of the CNN model, we now compare our proposed \textbf{Temperature-Aware} RFFP framework with other baselines. More specifically, the following baselines are compared:
\begin{itemize}[leftmargin=*]
    \item \textbf{Temperature-Unaware RFFP~\cite{nieves2026measurements, yilmaz2022effect}}: This is the most commonly utilized baseline by the literature, where power-normalized I/Q frames are used directly as an input, without including the temperature. 
    
    \item \textbf{Temperature-Invariant RFFP~\cite{yuan2025robust, shen2022towards}}: This approach utilizes the power-normalized I/Q frames as an input, but only after estimating and removing the residual Phase Offset (PO) and CFO---that are known to be highly impacted by hardware temperature variations. 
    
\end{itemize}

\begin{figure}
    \centering

    \setlength{\subfigcapskip}{-5pt}
    \includegraphics[width=0.85\linewidth]{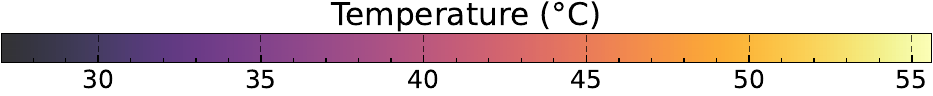}
    
    \setlength{\subfigcapskip}{-5pt}
    \includegraphics[width=\linewidth]{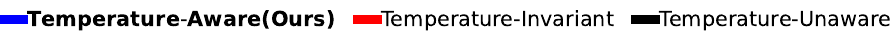}
    \vspace{-17.5pt}
    
    \subfigure[Tested on Loc1]{\includegraphics[width=0.325\linewidth]{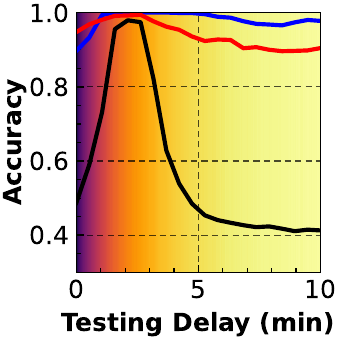}\label{subfig:AccuracyVsTime_Train_Round_Test_Round}}
    \subfigure[Tested on Loc2]{\includegraphics[width=0.325\linewidth]{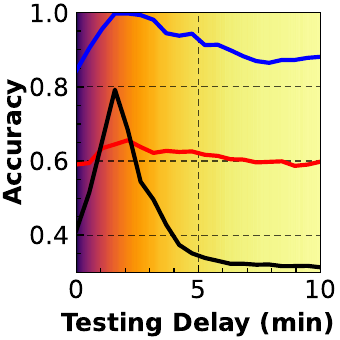}\label{subfig:AccuracyVsTime_Train_Round_Test_Round_NLoS}}    \subfigure[Tested on Loc3]{\includegraphics[width=0.325\linewidth]{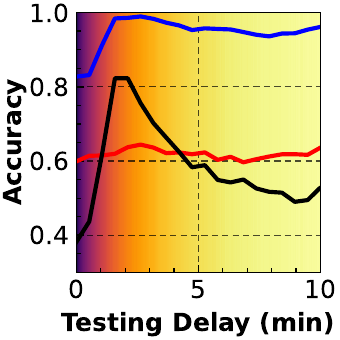}\label{subfig:AccuracyVsTime_Train_Round_Test_Table}}
    \caption{Accuracy vs. Time/Temperature when training at Loc1.}
    \label{fig:TrainLoc1}
\end{figure}

We mainly focused on the same experiment, where the training data is collected at temperature values between $30^\circ C$ to $52^\circ C$ (i.e., in the first 3 minutes from device power on) and the testing data is collected across the entire temperature range, including the unseen region of temperature values between $52^\circ C$ and $56^\circ C$ (i.e., from 3 minutes to 20 minutes from device power on). 

The accuracy results are presented in Figure~\ref{fig:TrainLoc1}, where training is performed using signals collected with the devices positioned at Loc1, while testing is conducted under each of the three considered locations: Loc1, Loc2, and Loc3. 
First, we want to mention that our results reveal that all methods exhibit relatively stable accuracy when evaluated on signals captured during the 10--20 minute interval. This behavior is expected because the device temperature remains approximately constant during that period, as previously shown in Figure~\ref{subfig:TempVsTime}. Therefore, the figures report only the results corresponding to the 0--10 minute interval, where the captured signals undergo significant temperature variation. 

Three key observations can be drawn from these results shown in Figure~\ref{fig:TrainLoc1}. First, the figure clearly demonstrates that our proposed framework outperforms the two baseline approaches in its ability to mitigate the effects of temperature variations. For example, Figure~\ref{fig:TrainLoc1}(a) shows that our Temperature-Aware approach maintains an accuracy above 97\% regardless of the testing data collection time, whereas the accuracies of the Temperature-Invariant and Temperature-Unaware approaches drop to approximately 90\% and 40\%, respectively, as the testing data collection time deviates from that of the training data collection time. 

Second, the superiority of our framework remains consistent across different locations, that is, even when the testing data is collected at a location (Figure~\ref{fig:TrainLoc1}(b) for Loc2 and Figure~\ref{fig:TrainLoc1}(c) for Loc3) that is different from that used during training, although all the accuracies obtained under each of the methods drop when compared to same-location testing data (Figure~\ref{fig:TrainLoc1}(a)).  

\begin{figure}
    \centering

    \setlength{\subfigcapskip}{-5pt}
    \includegraphics[width=0.85\linewidth]{figures/TempsLegend.pdf}
    
    \setlength{\subfigcapskip}{-5pt}
    \includegraphics[width=\linewidth]{figures/ResultsLegend.pdf}
    \vspace{-17.5pt}
    
    \subfigure[Tested on Loc1]{\includegraphics[width=0.325\linewidth]{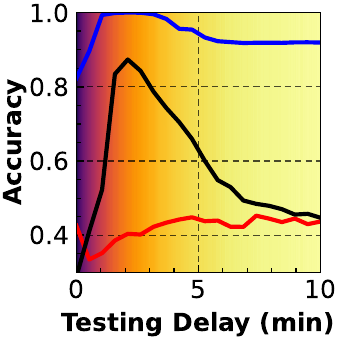}\label{subfig:AccuracyVsTime_Train_Round_NLoS_Test_Round}}
    \subfigure[Tested on Loc2]{\includegraphics[width=0.325\linewidth]{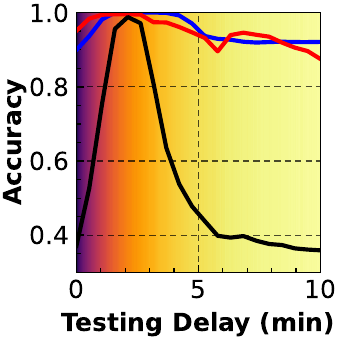}\label{subfig:AccuracyVsTime_Train_Round_NLoS_Test_Round_NLoS}}    \subfigure[Tested on Loc3]{\includegraphics[width=0.325\linewidth]{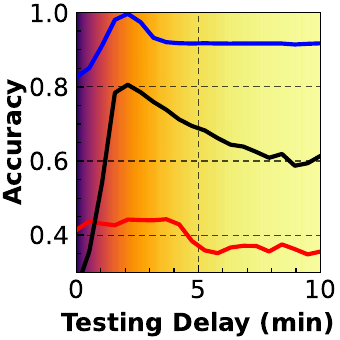}\label{subfig:AccuracyVsTime_Train_Round_NLoS_Test_Table}}
    \caption{Accuracy vs Time/Temperature when training at Loc2.}
    \label{fig:TrainLoc2}
\end{figure}

\begin{figure}
    \centering
    \setlength{\subfigcapskip}{-5pt}
    \includegraphics[width=0.85\linewidth]{figures/TempsLegend.pdf}
    
    \setlength{\subfigcapskip}{-5pt}
    \includegraphics[width=\linewidth]{figures/ResultsLegend.pdf}
    \vspace{-17.5pt}
    
    \subfigure[Tested on Loc1]{\includegraphics[width=0.325\linewidth]{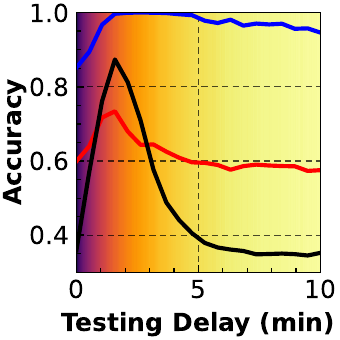}\label{subfig:AccuracyVsTime_Train_Table_Test_Round}}
    \subfigure[Tested on Loc2]{\includegraphics[width=0.325\linewidth]{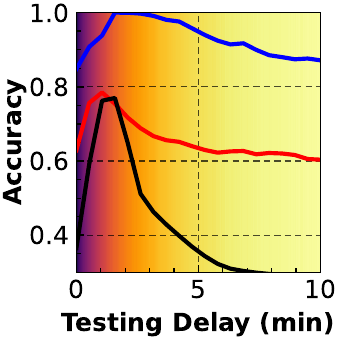}\label{subfig:AccuracyVsTime_Train_Table_Test_Round_NLoS}}    \subfigure[Tested on Loc3]{\includegraphics[width=0.325\linewidth]{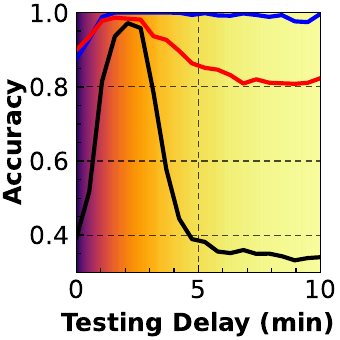}\label{subfig:AccuracyVsTime_Train_Table_Test_Table}}
    \caption{Accuracy vs Time/Temperature when training at Loc3.}
    \label{fig:TrainLoc3}
\end{figure}

Third, in the cross-location evaluation, the accuracies of the Temperature-Invariant and Temperature-Unaware approaches decrease much more significantly than that of the proposed framework. For example, Figure~\ref{fig:TrainLoc1}(b) shows that the accuracies of the Temperature-Invariant and Temperature-Unaware approaches drop from approximately 90\% and 40\% to around 60\% and 20\%, respectively. In contrast, the proposed framework experiences only a modest decline, from 97\% to approximately 92\%.
To evaluate the consistency of the proposed framework across different locations, Figures~\ref{fig:TrainLoc2} and~\ref{fig:TrainLoc3} present the testing accuracy when the training data is collected at Loc2 and Loc3, respectively. The results clearly demonstrate that the performance trends observed when the devices are placed at Loc1 remain consistent when changing the training locations.

% \vspace{-12pt}
\section{Conclusion}
\label{sec:conclusion}
In this paper, we investigated the impact of temperature variations on RFFP classification accuracy and demonstrated that ignoring temperature information can lead to significant performance degradation, particularly under unseen conditions. To address this challenge, we proposed a novel temperature-aware framework that incorporates real-time temperature measurements as additional input features.
Experimental results show that the proposed framework consistently improves classification accuracy and enhances generalization under both temperature and domain shifts. Furthermore, our findings indicate that temperature-aware modeling enables the classifier to more effectively capture the underlying hardware impairments dynamics, rather than relying solely on static features. 

% \section*{Acknowledgment}
% The preferred spelling of the word ``acknowledgment'' in America is without 
% an ``e'' after the ``g''. Avoid the stilted expression ``one of us (R. B. 
% G.) thanks $\ldots$''. Instead, try ``R. B. G. thanks$\ldots$''. Put sponsor 
% acknowledgments in the unnumbered footnote on the first page.

\bibliographystyle{IEEEtran}
\bibliography{References}

\end{document}